\documentclass[letter, reqno]{amsart}

\usepackage[margin=1.1in]{geometry}
\usepackage{amsmath}
\usepackage{xcolor}
\usepackage{setspace}
\usepackage{amsaddr}
\usepackage{natbib}
\usepackage{url}
\usepackage{bm}
\usepackage{float}
\usepackage{graphicx}
\usepackage{caption}
\usepackage{subcaption}

\newcommand{\E}{\mathbb{E}}

\newcommand{\X}{{\mathbf X}_{i}}
\newcommand{\Z}{Z_{i}}
\newcommand{\Y}{Y_{i}}
\def\Xv{\mathbf X}
\def\xv{\mathbf x}
\newtheorem{proposition}{Proposition}
\newtheorem{assumption}{Assumption}

\title[Differential recall bias in estimating treatment effects]{Differential recall bias in estimating treatment effects in observational studies}


\makeatletter
\g@addto@macro{\endabstract}{\@setabstract}
\newcommand{\authorfootnotes}{\renewcommand\thefootnote{\@fnsymbol\c@footnote}}%
\makeatother

\makeatletter
\def\blfootnote{\gdef\@thefnmark{}\@footnotetext}
\makeatother

\begin{document}
\setstretch{1.2}
\begin{center}
  \LARGE 
  \textbf{Differential Recall Bias in Estimating Treatment Effects \\in Observational Studies\blfootnote{This article has been accepted for publication in \textit{Biometrics} Published by Oxford University Press.}} \par \bigskip

  \normalsize
  \authorfootnotes
  Suhwan Bong\textsuperscript{1}, Kwonsang Lee\footnote{{Corresponding author}: \texttt{kwonsanglee@snu.ac.kr}}\textsuperscript{1} and Francesca Dominici\textsuperscript{2} \par \bigskip

  \textsuperscript{1}Department of Statistics, Seoul National University \par
  \textsuperscript{2}Department of Biostatistics, Harvard T.H. Chan School of Public Health\par \bigskip

\end{center}

\begin{abstract}
{
Observational studies are frequently used to estimate the effect of an exposure or treatment on an outcome. To obtain an unbiased estimate of the treatment effect, it is crucial to measure the exposure accurately. A common type of exposure misclassification is recall bias, which occurs in retrospective cohort studies when study subjects may inaccurately recall their past exposure. Particularly challenging is differential recall bias in the context of self-reported binary exposures, where the bias may be directional rather than random , and its extent varies according to the outcomes experienced. This paper makes several contributions: (1) it establishes bounds for the average treatment effect (ATE) even when a validation study is not available; (2) it proposes multiple estimation methods across various strategies predicated on different assumptions; and (3) it suggests a sensitivity analysis technique to assess the robustness of the causal conclusion, incorporating insights from prior research. The effectiveness of these methods is demonstrated through simulation studies that explore various model misspecification scenarios. These approaches are then applied to investigate the effect of childhood physical abuse on mental health in adulthood.
}
\end{abstract}

\setstretch{1.9}
\section{Introduction}
\label{sec1}
Observational studies are conducted to quantify the evidence of a potential causal relationship between an exposure or treatment and a given outcome. While numerous methods have been proposed to address confounding bias in observational studies, only a few have considered the challenges associated with accurately measuring exposure. One such challenge is the presence of \textit{recall bias}, which refers to a systematic error that occurs when participants inaccurately recall or omit details of past experiences, potentially influenced by subsequent events. Recall bias is particularly problematic in studies relying on self-reporting, such as retrospective cohort studies. It can lead to exposure misclassification, which can manifest as random or differential misclassification \citep{rothman2012epidemiology}. Unlike random recall bias, differential recall bias occurs when the misclassification of exposure information varies according to the value of other study variables. Specifically, under differential recall bias, exposure is differentially under-reported (over-reported) depending on the outcome. In addition to differential recall bias, random recall bias occurs when inaccuracies in the reporting of past events are due to chance and are not influenced by any specific factors. If the inaccuracies are equally likely to occur across the groups, then the bias may cancel out, and the estimated treatment effect may be unbiased \citep{raphael1987recall}. However, differential recall bias will likely lead to a biased estimate \citep{rothman2012epidemiology}. In retrospective cohort studies or case-control studies, differential recall bias cannot be eliminated even after adjusting for confounders. This paper will focus on methods for addressing differential recall bias in retrospective cohort studies.

In our motivating example, which examines the data on childhood physical abuse and adult anger, it is noted from prior research that significant differential recall bias can occur. This introduces a systematic bias, further compounding the bias already present due to confounders. Adults tend to under-report their exposures to childhood abuse because they are hesitant to disclose their experiences, even in anonymous or confidential surveys, due to feelings of shame, guilt, or fear of retaliation. Also, it is possible that individuals who have experienced childhood abuse and suffer from anger issues may be more likely to report their abuse, as their anger may be related to unresolved trauma or emotional distress stemming from the abuse. However, it is important to note that the relationship between childhood abuse and adult anger is complex, and under-reporting of childhood abuse is a common problem that can vary depending on a range of individual and contextual factors \citep{fergusson2000stability}.

There is an extensive literature on measurement error correction and exposure misclassification in epidemiology \citep{carroll1995measurement, rothman2008modern}. For instance, numerous studies have concentrated on the error-in-variables model, addressing measurement error in linear regression problems \citep{lindley1953estimation, lord1960large, cochran1968errors, fuller1980properties, carroll1985comparison}. The regression calibration algorithm is proposed as a general approach by \cite{carroll1990approximate} and \cite{gleser1990improvements}. Several measurement error correction methods have also been developed for binary misclassification problems and logistic regression models \citep{bross1954misclassification, armstrong1985measurement, stefanski1985covariate, rosner1989correction, rosner1990correction}. However, previous methodologies addressing measurement error have predominantly focused on outcome prediction and targeted the association of variables. In our research, we aim to address this issue while considering confounders and focusing on the causal relationship between variables. To the best of our knowledge, contributions regarding the impact of differential recall bias on measures with causal interpretations are scarce.

Accounting for measurement error in causal inference is important. However, studies on measurement error have typically focused on mismeasured covariates and misclassified outcomes. For example, previous studies have investigated measurement error in covariates \citep{mccaffrey2013inverse, lockwood2016matching} and misclassification of binary outcomes \citep{gravel2018weighted}. However, only a limited number of studies have addressed misclassified exposure, or recall bias. \cite{imai2010causal} proposed a nonparametric identification method for estimating the average treatment effect (ATE) with differential treatment measurement error. This method could address both over-reporting and under-reporting measurement errors, but it was based on strict assumptions, such as no misclassification for compliant groups. Furthermore, their bounds for the ATE were derived from the true treatment assigning probability, which may be unknown when recall bias is present. \cite{babanezhad2010comparison} and \cite{braun2017propensity} have shown that the exposure misclassification could significantly impact causal analysis. \cite{babanezhad2010comparison} compared several causal estimators for time-varying exposure reclassification cases, and \cite{braun2017propensity} proposed a likelihood-based method that adjusts for exposure misclassification bias, presupposing on non-differential measurement error assumption. In summary, these studies highlight a gap regarding causal inference methodologies that can analytically assess the impact of differential recall bias.

The primary goal of this paper is to introduce a collection of robust estimators for estimating the ATE in the presence of differential recall bias. We emphasize the significance of delineating the impact of differential recall bias on ATE within the causal inference framework. Additionally, this paper has 3 further contributions. First, we derive bounds for the ATE that do not rely on a validation study. The bounds can be refined with additional information about the nature of differential recall bias. This allows researchers to tailor the bounds based on the available evidence. Second, we propose multiple estimation methods using 2 different strategies for estimating the ATE—maximum likelihood estimation (MLE) and stratification. Each of the methods is based on its own assumptions, offering valuable insights on potential model misspecification. Finally, we propose a novel sensitivity analysis approach to assess the impact of differential recall bias on our conclusion. This is a crucial and useful way to quantify the evidence, given that the degree of differential recall bias is typically unknown in practice. We illustrate the application and efficacy of this sensitivity analysis method by applying it to the real data in Section~\ref{sec6}.

\section{Notation and Recall Bias Model}
\label{sec2}

\subsection{Causal Inference Framework and Target Parameters}
\label{sec2.1}

We start by introducing the potential outcome framework \citep{rubin1974estimating}. Assume $N$ individuals in total. We denote $\X \in \mathcal{X} \subseteq \mathbb{R}^d$ as an observed covariate vector for the $i$th individual. We let $\Z=1$, to indicate that individual $i$ was exposed to a certain binary exposure, and $\Z=0$ otherwise. We can define potential outcomes as follows: If $\Z=0$, then individual $i$ exhibits response $\Y(0)$; if $\Z=1$, then individual $i$ exhibits $\Y(1)$. Only one of the two potential outcomes can be observed depending on the exposure of individual $i$. The response exhibited by individual $i$ is $\Y=\Z\Y(1)+(1-\Z)\Y(0)$. In this paper, $\Y(0)$ and $\Y(1)$ are assumed to be binary. Depending on the occurrence of the outcome, the potential outcome is equal to either 0 or 1. We consider two assumptions: (1) unconfoundedness and (2) positivity. The unconfoundedness assumption means that the potential outcomes $(\Y(0), \Y(1))$ are conditionally independent of the treatment $Z$ given $\X$, that is, $(\Y(0),\Y(1)) \perp\!\!\!\perp \Z\vert\X$. The positivity assumption means that the probability $\Pr(\Z=1\vert\X)$ lies in $(0,1)$. These assumptions together are often called \textit{strong ignorability} \citep{rosenbaum1983central}. We also adopt the Stable Unit Treatment Value Assumption \citep{rubin1980randomization} to identify causal effects; that is, the potential outcomes for each individual are not affected by the treatment status of other individuals.

Binary exposure is frequently retrospectively investigated to find the cause of the outcome in observational studies; thus, exposure to a risk factor is never randomized. A naive comparison of the prevalence of the outcome between the exposed and unexposed groups can be misleading due to the confounding bias. The effect caused by treatment to an individual $i$ is defined as the difference, $\Y(1)-\Y(0)$. However, it is impossible to observe both $\Y(0)$ and $\Y(1)$ for any individual. Under strong ignorability assumptions, it is possible to identify the ATE. Thus, our parameter of interest is the ATE, $\tau = \E[\Y(1)] - \E[\Y(0)]$. In some instances, we are interested in estimating the \textit{conditional average treatment effect} (CATE) at a given level of $\X=\xv$ for $\xv\in\mathcal{X}$ as $\tau(\xv) = \E[\Y(1)\vert\X=\xv] - \E[\Y(0)\vert\X=\xv]$.

\subsection{Differential Recall Bias Model}
\label{sec2.2}
Some observational studies, including retrospective cohort studies, are retrospective in nature. Thus, recall bias may occur when the exposures are self-reported. In this paper, we consider situations with differential recall bias where the exposure is under-reported (over-reported) differently depending on the outcome. In the presence of differential recall bias, the underlying true exposure $\Z$ is not observed. Instead, we observe the biased exposure $\Z^*$ with recall bias. If no recall bias exists, then $\Z=\Z^*$. In the childhood abuse example, previous literature has indicated that exposure was mainly under-reported. To address this issue, we assume that either over-reporting or under-reporting recall biases occur. Since over-reporting can be treated similarly, we focus on the under-reporting recall bias in this paper.

\begin{assumption}[Differential Recall Bias] \label{assumption1}
    Recall bias occurs independently with probability $\eta_y(\xv)$ for individuals with $\Y=y$, $\Z=1$, and $\X=\xv$ where $y = 0, 1$ and $\xv\in\mathcal{X}$.
    \begin{align*}
        \eta_0(\xv) = \Pr(\Z^*=0\vert \Y=0,\Z=1,\X=\xv),\\
        \eta_1(\xv) = \Pr(\Z^*=0\vert \Y=1,\Z=1,\X=\xv).
    \end{align*}
\end{assumption}

Assumption~\ref{assumption1} proposes the differential recall bias model that assumes that the occurrence and magnitude of bias depend on the observed outcome and $\X$. In essence, after stratifying the data based on $\X$, the $2 \times 2$ contingency table of $Y_i$ and $Z_i$ can be subject to misclassification. The parameters $\eta_0(\xv)$ and $\eta_1(\xv)$ represent the probability of under-reporting depending on outcome variables. In under-reported cases, $\Z=0$ always implies $\Z^*=0$, but $\Z=1$ implies either $\Z^*=1$ or $\Z^*=0$. Therefore, recall bias occurs only when $\Z=1$. Note that if there is no recall bias, then $\eta_0(\xv)=\eta_1(\xv)=0$.

\section{Identification of Causal Parameters}
\label{sec3}
If recall bias is absent and the exposure is observed correctly, then $\tau$ and $\tau(\xv)$ can be identified under strong ignorability assumptions. However, in the presence of recall bias, if the inference is made on the basis of observed  $\Z^*$ rather than $\Z$, then we obtain a biased estimate because $(\Y(0),\Y(1)) \not\perp\!\!\!\perp \Z^* \vert \X$. To describe this, consider the probabilities based on observed exposure $Z^\ast$, $p_{y\vert z}^\ast(\xv) = \Pr(\Y=y\vert \Z^\ast=z,\X=\xv)$ for $z=0,1$, $y=0,1$, and $\xv\in\mathcal{X}$. Then,
\begin{align*}
    p_{y\vert z}(\xv) = \frac{p_{yz}(\xv)}{p_{1z}(\xv)+p_{0z}(\xv)} \not= \frac{p_{yz}^\ast(\xv)}{p_{1z}^\ast(\xv)+ p_{0z}^\ast(\xv)} = p_{y\vert z}^\ast(\xv)
\end{align*}
where $p_{yz}(\xv) = \Pr(\Y=y,\Z=z\vert \X=\xv)$ and $p_{yz}^\ast(\xv) = \Pr(\Y=y,\Z^\ast=z\vert \X=\xv)$ for $z=0,1$, $y=0,1$, and $\xv\in\mathcal{X}$ holds.

The precise recall bias mechanism in real-life scenarios is often unknown. Assuming we lack precise knowledge of the recall bias parameter functions stated in Assumption~\ref{assumption1}, it becomes infeasible to identify the ATE with certainty. However, if we can establish bounds on the recall bias parameter functions, we can potentially bound the target parameters. Under the recall bias model in Assumption~\ref{assumption1}, the following relationships hold:
\begin{align} \label{relationship1}
    p_{11}(\xv) = \frac{p_{11}^\ast(\xv)}{1-\eta_1(\xv)}, \quad
    p_{10}(\xv) = p_{10}^\ast(\xv) - \frac{\eta_1(\xv)}{1-\eta_1(\xv)} p_{11}^\ast(\xv), \nonumber\\
    p_{01}(\xv) = \frac{p_{01}^\ast(\xv)}{1-\eta_0(\xv)}, \quad
    p_{00}(\xv) = p_{00}^\ast(\xv) - \frac{\eta_0(\xv)}{1-\eta_0(\xv)} p_{01}^\ast(\xv).
\end{align}

The following proposition allows partial identification of the causal parameters if the recall bias occurs with the probability of at most $\delta$.

\begin{proposition}\label{proposition1}
    Under Assumption~\ref{assumption1}, suppose there exists a constant $0\leq\delta<1$ which $0\leq\eta_0(\xv),\eta_1(\xv)\leq\delta$ holds for all $\xv\in\mathcal{X}$. Then the following inequalities hold for all $\xv\in\mathcal{X}$:
    \begin{align*}
        \frac{p_{11}^*(\xv)}{p_{11}^*(\xv)+\frac{1}{1-\delta}p_{01}^*(\xv)} &\leq p_{1|1}(\xv) \leq \frac{p_{11}^*(\xv)}{p_{11}^*(\xv)+(1-\delta)p_{01}^*(\xv)}, \nonumber \\
        \frac{p_{10}^*(\xv)-\frac{\delta}{1-\delta}p_{11}^*(\xv)}{p_{10}^*(\xv)+p_{00}^*(\xv)-\frac{\delta}{1-\delta}p_{11}^*(\xv)} &\leq p_{1|0}(\xv) \leq \frac{p_{10}^*(\xv)}{p_{10}^*(\xv)+p_{00}^*(\xv)-\frac{\delta}{1-\delta}p_{01}^*(\xv)}.
    \end{align*}
\end{proposition}

This proposition can be used when we can constrain the occurrence probability of recall bias using domain knowledge.

Some additional assumptions may be useful to narrow the bounds of estimands. For example, when studying the potential impact of childhood abuse on mental health issues in adulthood, it is important to consider the possibility of individuals hiding or feeling shame about their previous experiences. Additionally, those who have mental health problems in adulthood may be more likely to under-report their history of abuse, as they may feel particularly affected by their experiences and may be hesitant to disclose them. Providing this additional information enables us to make the assumption that $\eta_0(\xv) \leq \eta_1(\xv)$.

\clearpage
\begin{proposition}\label{proposition2}
    Under Assumption~\ref{assumption1},
    \begin{enumerate}
        \item[(a)] Suppose there exists a constant $0\leq\delta<1$ which $0\leq\eta_0(\xv)\leq \eta_1(\xv)\leq\delta$ holds for all $\xv\in\mathcal{X}$. Then, the following inequalities hold for all $\xv\in\mathcal{X}$:
        \begin{gather*}
            p^*_{1|1}(\xv) \leq p_{1|1}(\xv) \leq \frac{p_{11}^*(\xv)}{p_{11}^*(\xv)+p_{01}^*(\xv)(1-\delta)}, \\
            \frac{p_{10}^*(\xv)-\frac{\delta}{1-\delta}p_{11}^*(\xv)}{p_{10}^*(\xv)+p_{00}^*(\xv)-\frac{\delta}{1-\delta}p_{11}^*(\xv)} \leq p_{1|0}(\xv) \leq \max\left\{p^*_{1|0}(\xv), \frac{p_{10}^*(\xv)-\frac{\delta}{1-\delta}p_{11}^*(\xv)}{p_{10}^*(\xv)+p_{00}^*(\xv)-\frac{\delta}{1-\delta}\{p_{01}^*(\xv)+p_{11}^*(\xv)\}}\right\}.
        \end{gather*}
        \item[(b)] Suppose there exists a constant $0\leq\delta<1$ which $0\leq\eta_1(\xv)\leq \eta_0(\xv)\leq\delta$ holds for all $\xv\in\mathcal{X}$. Then, the following inequalities hold for all $\xv\in\mathcal{X}$:
        \begin{gather*}
            \frac{p_{11}^*(\xv)}{p_{11}^*(\xv)+p_{01}^*(\xv)/(1-\delta)}\leq p_{1|1}(\xv) \leq p^*_{1|1}(\xv), \\
            \min\left\{p^*_{1|0}(\xv), \frac{p_{10}^*(\xv)-\frac{\delta}{1-\delta}p_{11}^*(\xv)}{p_{10}^*(\xv)+p_{00}^*(\xv)-\frac{\delta}{1-\delta}\{p_{01}^*(\xv)+p_{11}^*(\xv)\}}\right\} \leq p_{1|0}(\xv) \leq \frac{p_{10}^*(\xv)}{p_{10}^*(\xv)+p_{00}^*(\xv)-\frac{\delta}{1-\delta}p_{01}^*(\xv)}.
        \end{gather*}
    \end{enumerate}
\end{proposition}

This proposition implies that by assuming a relationship between two parameters, $\eta_0(\xv)$ and $\eta_1(\xv)$, we can narrow down either the upper or lower bound of the ATE. We can partially identify the causal parameter when the exact recall bias parameter functions are unknown. We can also point identify the causal treatment effect parameters if the recall bias parameter functions are known.

\begin{proposition}\label{proposition3}
    Under Assumption~\ref{assumption1}, the following equality holds for all $\xv\in\mathcal{X}$.
    \begin{align*}
        \tau(\xv) = \frac{\frac{p_{11}^*(\xv)}{1-\eta_1(\xv)}}{\frac{p_{11}^*(\xv)}{1-\eta_1(\xv)} +\frac{p_{01}^*(\xv)}{1-\eta_0(\xv)}} - \frac{p_{10}^*(\xv) - \frac{\eta_1(\xv)}{1-\eta_1(\xv)} p_{11}^*(\xv)}{p_{10}^*(\xv) - \frac{\eta_1(\xv)}{1-\eta_1(\xv)} p_{11}^*(\xv)+p_{00}^*(\xv) - \frac{\eta_0(\xv)}{1-\eta_0(\xv)} p_{01}^*(\xv)}.
    \end{align*}
\end{proposition}
\medskip

If a validation study is available, $\eta_0(\xv), \eta_1(\xv)$ can be estimated and then be plugged into the above equation. However, it is not common in many situations, especially at an early stage of research.

\clearpage
\section{Methods for Recovering the Treatment Effects in the Presence of Recall Bias}
\label{sec4}
In this section, we propose two estimation methods that provide consistent estimates of the ATE in the presence of recall bias and confounding: (1) maximum likelihood estimation and (2) stratification. We suggest three stratification techniques for the stratification method: (1) propensity score stratification, (2) prognostic score stratification, and (3) blocking. Furthermore, we discuss the nearest-neighbor combination method used to address the problems in the stratification method with recall bias.

For given $\eta_0(\xv)$ and $\eta_1(\xv)$, the first maximum likelihood (ML)-based method requires the correct identification of models for exposure and two potential outcomes to obtain a consistent estimate of $\tau$. The stratification-based method requires a few model assumptions. Stratification can be implemented on the basis of either propensity scores or prognostic scores \citep{hansen2008prognostic}. The propensity score stratification method requires a correctly specified exposure model, while the prognostic score stratification method needs a correctly specified outcome model. The last blocking method suggested by \cite{karmakar2021reinforced} does not need any model assumption. In the following subsections, we discuss these estimation methods in more detail.

\subsection{Maximum Likelihood Estimation}
\label{sec4.1}
Consider the outcome models $m_z(\xv) = \Pr(\Y=1\vert\Z=z,\X=\xv)$ for $z=0,1$ that are the models for the two probabilities, $p_{1\vert1}(\xv)$ and $p_{1\vert0}(\xv)$and, the propensity score model $e(\xv)$ for $\Pr(\Z=1\vert\X=\xv)$. The probability $p_z(\xv) = \E[\Y(z) | \X = \Xv]$ can be identified as $p_{1\vert z}(\xv)$, which can thus be estimated by $m_1(\xv)$. It is well-known that, in the absence of recall bias, either $m_z(\xv)$ or $e(\xv)$ is required to be correctly specified to obtain a consistent estimate. However, in the presence of recall bias, $m_z(\xv)$ nor $e(\xv)$ cannot be estimated from the observable data set due to the absence of true $\Z$. We can rather estimate the ATE as a function of the tuning parameters of the recall bias model.

The first method presented in this subsection uses maximum likelihood estimation. $m_z(\xv)$ and $e(\xv)$ must be specified to construct the likelihood function to obtain an estimate for given $\eta_0(\xv)$ and $\eta_1(\xv)$. Under Assumptions~\ref{assumption1}, the joint probability $\Pr(\Y,\Z^\ast\vert\X)$ of observable variables can be represented by a function of $m_0(\xv)$, $m_1(\xv)$, and $e(\xv)$. We assume models $m_z(\xv; \bm\gamma_z), z=0, 1$ and $e(\xv; \bm\beta)$ with parameters $\bm\gamma_z$ and $\bm\beta$, respectively. For instance, logistic regressions can be used such as $m(Z, \Xv; \bm\gamma) = \exp(\gamma_z Z + \bm\gamma_{\Xv}^T \Xv)/\{1+ \exp(\gamma_z Z + \bm\gamma_{\Xv}^T \Xv )\}$ with $m_1(\Xv) = m(1, \Xv; \bm\gamma)$ and $m_0(\Xv) = m(0, \Xv; \bm\gamma)$ and $e(\Xv) = \exp({\bm\beta}^T \Xv) /\{1 + \exp(\bm\beta^T \Xv)\}$. These model parameters can be estimated by solving the following maximization problem: 
\begin{equation*} \label{mle}
	\widehat{\bm\theta} = (\widehat{\bm\beta}, \widehat{\bm\gamma}_0, \widehat{\bm\gamma}_1) = \operatorname*{argmax}_{\bm\beta, \bm\gamma_0, \bm\gamma_1} \sum_{i=1}^{N} \log \Pr(Y_{i} = y_{i}, Z_{i}^* = z_{i} | \Xv_{i} = \xv_{i}).
\end{equation*}

Once we obtain the estimate $\widehat{\bm \theta}$, we can compute $\widehat{m}_z(\xv) = m_t(\xv; \widehat{\bm\gamma}_z)$ and $\widehat{e}(\xv) = e(\xv; \widehat{\bm\beta})$. The marginal probabilities ${p}_z$ are then estimated by taking sample averages of $\widehat{m}_z(\xv)$ as $\widehat{p}_{1}^{ML} = \frac{1}{N} \sum_{i=1}^{N} \widehat{m}_1(\X)$ and $\widehat{p}_{0}^{ML} = \frac{1}{N} \sum_{i=1}^{N} \widehat{m}_0(\X)$. Then, the ATE estimate is $ \widehat{\tau}^{ML} = \widehat{p}_{1}^{ML} - \widehat{p}_{0}^{ML}$. This estimate is consistent if $m_z(\xv; \bm\gamma_z)$ and $e(\xv; \bm\beta)$ must be correctly specified for fixed $\eta_0(\xv), \eta_1(\xv)$.

A key challenge with this method is the requirement for researchers to specify $\eta_0(\xv)$ and $\eta_1(\xv)$, which are typically unknown in practice. If these can be estimated from external sources, such estimates can be incorporated into the likelihood function. In the absence of information about $\eta_0(\xv)$ and $\eta_1(\xv)$, an additional assumption might be made that $\eta_0(\xv) = \eta_0$ and $\eta_1(\xv) = \eta_1$, implying constant values across all $\xv$. To address this uncertainty, a sensitivity analysis could be employed, utilizing a plausible range for $\eta_0$ and $\eta_1$ informed by prior research. This involves testing every possible combination of $(\eta_0, \eta_1)$ within the specified range and assessing the impact on the ATE estimate's variation.

\subsection{Stratification} \label{sec4.2}
Stratification can be alternatively used to estimate $\tau$ by aiming to balance the covariate distributions between exposed and unexposed groups. Compared with the ML method, stratification requires fewer assumptions in general. If stratification can be successfully created while adjusting for confounders, the estimation of $\tau$ is straightforward. Assume that there are $I$ strata. Each stratum $i$, contains $n_i$ individuals. There are $N = \sum_{i=1}^I n_i$ individuals in total. Denote $ij$ as the $j$th individual in stratum $i$ for $j=1,\dots, n_i$. If we assume $(Y_{ij}(1), Y_{ij}(0))\perp Z_{ij}$ holds within each stratum $i$, then the stratum-specific probabilities $p_{1i}=\E_{\Xv\vert\textbf{stratum i}}[p_1(\Xv)]$ and $p_{0i}=\E_{\Xv\vert\textbf{stratum i}}[p_0(\Xv)]$ can be identified from the $2\times2$ table generated by stratum $i$. However, $Z_{ij}^\ast$ is observed instead of $Z_{ij}$ due to recall bias. Therefore, the recall bias adjustment using \eqref{relationship1} is required. For stratum $i$, assume that Table~\ref{table1} with $a_i^*, b_i^*, c_i^*, d_i^*$ is observed.

\clearpage

\begin{table}[ht]
	\centering
	\caption{The $2 \times 2$ contingency observed table for the $i$th stratum.}
	\begin{tabular}{c|cc|c}
		&  $Y=1$ & $Y=0$ & \\
		\hline
		Exposed $(Z^*=1)$ & $a_i^*$ & $b_i^*$ & $a_i^*+b_i^*$ \\
		Not exposed $(Z^*= 0)$ & $c_i^*$ & $d_i^*$ & $c_i^*+d_i^*$ \\
		\hline
		& $a_i^* + c_i^*$ & $b_i^*+d_i^*$ & $n_i^*$
	\end{tabular}
	\label{table1}
\end{table}

\begin{proposition}\label{proposition4}
    Suppose there are $2 \times 2$ contingency tables for $I$ strata on $Y$ and $Z^*$ with $a_i^*, b_i^*, c_i^*, d_i^*$ as in Table~\ref{table1}. The stratum-specific ATE $\tau_i = \E_{\Xv\vert\textbf{stratum i}}[\tau(\xv)]$ can be estimated for known $\eta_0(\xv), \eta_1(\xv)$ as
    \begin{align*}
    &\widehat{\tau}_i = \frac{\sum_{j=1}^{n_i} \frac{Z_{ij}^* Y_{ij}}{1-\eta_1(\xv_{ij})} }{\sum_{j=1}^{n_i} \frac{Z_{ij}^* Y_{ij}}{1-\eta_1(\xv_{ij})} + \sum_{j=1}^{n_i} \frac{Z_{ij}^*(1-Y_{ij})}{1-\eta_0(\xv_{ij})} } \\
    &- \frac{\sum_{j=1}^{n_i}(1-Z_{ij}^*) Y_{ij} - \sum_{j=1}^{n_i} Z_{ij}^* Y_{ij}\frac{\eta_1(\xv_{ij})}{1-\eta_1(\xv_{ij})} }{ \sum_{j=1}^{n_i}(1-Z_{ij}^*) Y_{ij} - \sum_{j=1}^{n_i} Z_{ij}^* Y_{ij}\frac{\eta_1(\xv_{ij})}{1-\eta_1(\xv_{ij})} + \sum_{j=1}^{n_i}(1-Z_{ij}^*) (1-Y_{ij}) - \sum_{j=1}^{n_i} Z_{ij}^*(1-Y_{ij})\frac{\eta_1(\xv_{ij})}{1-\eta_1(\xv_{ij})}}.
    \end{align*}

Also, for $0 \leq \eta_0(\xv), \eta_1(\xv) \leq \delta_i$, the bound can be estimated as
$$
\frac{a_i^*}{a_i^* + b_i^*\frac{1}{1-\delta_i}} - \frac{c_i^*}{c_i^* + d_i^* - b_i^* \frac{\delta_i}{1-\delta_i} } \leq \widehat{\tau}_i \leq \frac{a_i^*}{a_i^* + b_i^*(1-\delta_i)} - \frac{c_i^* - a_i^*\frac{\delta_i}{1-\delta_i}}{c_i^* + d_i^*- a_i^*\frac{\delta_i}{1-\delta_i}}.
$$

\end{proposition}

This proposition is directly obtained from Propositions~\ref{proposition1} and \ref{proposition3}. If $\eta_0(\xv) = \eta_0$ and $\eta_1(\xv) = \eta_1$, the estimate $\widehat{\tau}_i$ is simplified as $\frac{a_i^*/(1-\eta_1)}{a_i^*/(1-\eta_1) + b_i^*/(1-\eta_0)} - \frac{c_i^* - a_i^*(\eta_1/(1-\eta_1))}{c_i^* - a_i^*(\eta_1/(1-\eta_1)) + d_i^* - b_i^*(\eta_0/(1-\eta_0))}$. The marginal probabilities can be estimated by the weighted average of these stratum-specific $\widehat{\tau}_i$ with weights $s_i = n_i/N$. Therefore, the ATE is estimated by $\widehat{\tau}^{S} = \sum_{i=1}^{I} \widehat{\tau}_i (s_i/N)$. The bound can be similarly obtained. 

\subsubsection{Propensity Score Stratification}
\label{sec4.2.1}
Among stratification-based methods, stratification based on propensity score is the most common approach \citep{rosenbaum1983central}. The propensity score is a conditional probability of the treatment assignment given the observed covariates, $e(\xv) = \Pr(\Z=1\vert \Xv_i=\xv)$. We only have to assume a treatment model to create strata using propensity scores. However, similar to many stratification-based methods, this method relies on the assumption that stratification achieves covariate balance by at least approximately. Furthermore, strata are formed on the basis of biasedly estimated propensity score $\widehat{e}^*(\xv) = \Pr(\Z^*=1\vert \Xv_i = \xv)$ using $Z^*$ instead of unobservable $Z$. It is not feasible to compare the covariate distributions between the exposed and unexposed groups. Thus, constructing strata based on the propensity score can be problematic if $\eta_0(\xv)$ and $\eta_1(\xv)$ are significantly different from $0$. However, if $\eta_0(\xv) = \eta_1(\xv) = \eta$ for all $\xv\in\mathcal{X}$, then the covariate balance between the $Z^*=1$ and $Z^*=0$ groups is asymptotically the same as that between the $Z=1$ and $Z=0$ groups since $e^\ast(\xv) = (1-\eta)e(\xv)$. If the recall bias occurs with the same probability across the $Y=0$ and $Y=1$ groups (i.e., recall bias is not differential), then $e^\ast(\xv)$ is also a balancing score. Thus, we can create valid strata using the biased propensity score obtained by observable variables.

\subsubsection{Prognostic Score Stratification}
\label{sec4.2.2}
Instead of using the propensity score, the prognostic score can be utilized to construct strata \citep{hansen2008prognostic}. If there is $\Psi(\Xv_{ij})$ such that $Y_{ij}(0) \perp\!\!\!\perp \Xv_{ij} \vert \Psi(\Xv_{ij})$, then we call $\Psi(\cdot)$ the prognostic score. Similar to propensity score stratification, prognostic score stratification permits the estimation of exposure effects within the exposed group. If $(Y_{ij}(0), Y_{ij}(1)) \perp\!\!\!\perp \Xv_{ij} \vert \Psi(\Xv_{ij})$ is further assumed, then prognostic score stratification is valid for estimating overall exposure effects. For instance, if  $m(Z_{ij}, \Xv_{ij}; \bm\gamma) = \exp(\gamma_z Z_{ij} + \bm\gamma_{\Xv}^T \Xv_{ij})/\{1+ \exp(\gamma_z Z + \bm\gamma_{\Xv}^T \Xv_{ij} )\}$ is assumed, $\Psi(\Xv_{ij}) = \bm\gamma_{\Xv}^T \Xv_{ij}$ is the prognostic score.

Like propensity score stratification, stratification on the prognostic score leads to a desirable and balanced structure. Since we do not know $\Psi(\Xv_{ij})$ a priori, it has to be estimated from the data. As mentioned before, if $\eta_0(\xv) = \eta_1(\xv) = \eta$, then the probabilities of recall bias occurrence between the $Y=1$ and $Y=0$ groups are the same. In this case, the prognostic score can be used in stratification while estimating the treatment effect. Since the exposure was under-reported, we know $Z_{ij}^* = 1$ always implies $Z_{ij} = 1$. We first estimate $\gamma_{\Xv}$ by using the data of the $\Z^* = 1$ group. Assuming that the recall bias occurs randomly, we then calculate the prognostic scores $\Psi(\Xv_{ij}) = \bm\gamma_{\Xv}^T \Xv_{ij}$ for all individuals. The outcome models should be correctly specified for prognostic score stratification. Even though $\widehat\tau^{Prog}$ needs fewer modeling assumptions than $\widehat\tau^{ML}$, modeling assumption is still required. Moreover, score-based stratifications need a further assumption that $\eta_0(\xv) = \eta_1(\xv) = \eta$ to be justified.

\subsubsection{Blocking}
\label{sec4.2.3}
In Sections~\ref{sec4.2.1} and \ref{sec4.2.2}, proper scores based on modeling assumptions are required to create valid strata. Also, score-based stratifications could be problematic if $\eta_0(\xv)$ and $\eta_1(\xv)$ significantly differ. Stratification based on propensity score also requires accurate treatment model identification, and the outcome model must be correctly specified to create strata with a prognostic score. However, the blocking method does not require any model assumption. Our goal is to make covariates $\Xv_{ij}$ in block $i$ to be similar. If the covariates in each block are almost the same, then we assume that $(Y_{ij}(0),Y_{ij}(1))\perp Z_{ij}$ in each block $i$ holds. \cite{karmakar2021reinforced} used the \textit{blockingChallenge} package in R to build blocks.

Suppose that there are $N=Ik$ individuals. To make $I$ blocks with size $k$, $I$ individuals are first randomly chosen as template individuals for each block. The remaining $I(k-1)$ individuals are then matched to template individuals using optimal matching at a ratio of $(k-1):1$. After the first blocking, separate an individual who is the most distant from the remaining $k-1$ individuals in each block. Setting these $I$ individuals as template individuals for each block, optimal matching is used again to build $I$ blocks. Repeating this process facilitates the implementation of an effective minimum within-block distance stratification. Repeat this process until no changes occur to obtain $I$ blocks, which are strata with size $k$.

The blocking method does not require any model assumption. However, the covariates $\Xv_{ij}$ in each block $i$ need to be similar. When achieving covariate balance is difficult or a weak overlap situation emerges, such blocks are not obtained. If the covariate balance within the block can be easily achieved, the blocking method is likely to provide a reliable estimator. Different from $\widehat\tau^{ML}$, $\widehat\tau^{Prop}$, and $\widehat\tau^{Prog}$, an advantage of $\widehat\tau^{B}$ is that any modeling assumptions is unnecessary. This stratification technique is still robust under model misspecification. We will examine the performances of these estimators in Section~\ref{sec5}.

\section{Simulation Studies}
\label{sec5}
We conduct simulation studies to compare the performance of the proposed methods: (1) ML, (2) propensity score stratification, (3) prognostic score stratification, and (4) blocking. We consider various model specification scenarios to examine how they can successfully recover the true treatment effect under different model misspecification cases. In addition, we include Naïve estimators based on inverse probability weighting (IPW) and outcome regression (OR), assuming no misclassification error.

We consider four independent covariates, $\Xv_{i}=(X_{i1}, X_{i2}, X_{i3}, X_{i4})$. $X_{i1}$ and $X_{i2}$ are binary covariates, whereas $X_{i3}$ and $X_{i4}$ are continuous covariates. We also consider four simulation scenarios where the exposure and outcome models are correctly specified or misspecified: (i) (cor, cor), (ii) (cor, mis), (iii) (mis, cor), and (iv) (mis, mis). For example, (mis, cor) means the exposure model is misspecified, but the outcome model is correctly specified. We randomly generate exposure $\Z$ and potential outcomes $(\Y(0), \Y(1))$ of each individual depending on the model specification scenario. However, due to recall bias, we cannot observe the true exposure $\Z$; we observe the biased exposure $\Z^\ast$ instead. We assume that the exposure is under-reported for this simulation study. We generate $\Z^\ast$ based on the observed outcome $\Y = \Y(1)\Z + \Y(0)(1-\Z)$ (See Web Appendix for the detailed simulation settings).

We compare the considered methods considering their successful recovery of the true ATE under different model misspecification scenarios. In addition to this factor of model misspecification, we also consider two sample sizes ($N=1000$ or $2000$) and two constant recall bias parameter functions ($(\eta_0(\xv), \eta_1(\xv)) = (0.1,0.1)$ or $(0.1,0.2)$) throughout this simulation. We fix the strata size to 50 in stratification methods with the nearest-neighbor combination method. Table~\ref{table2} shows the simulation results that are obtained from 1000 simulated datasets.

Naïve estimators assuming no misclassification error exhibit poor performance across various model misspecification scenarios, particularly in cases with a differential recall bias. This highlights the necessity of adjusting bias when we overlook the potential for exposure misclassification. Among the estimators we proposed, if both the exposure and potential outcome models are correctly specified, then $\widehat\tau^{ML}$ is the best estimator. $\widehat\tau^{Prop}$ and $\widehat\tau^{Prog}$ show similar performance in each scenario. Particularly, even in the treatment model misspecification scenario, stratification based on propensity score shows slightly better results than stratification based on prognostic score. Score-based stratifications perform agreeably, although $\eta_0$ and $\eta_1$ are different. $\widehat\tau^{B}$ provides the least biased estimate in the case of misspecification for both models. On the contrary, $\widehat\tau^{ML}$ shows the worst performance in (mis, mis) scenario. As expected, the model dependency for the blocking method is the smallest, and that for the ML method is the largest. This finding leads to a good result of the blocking estimator and a poor result of the ML estimator in the worst model misspecification scenario. Even though we require weak assumptions, the blocking estimator performs well throughout every model misspecification scenario. If the models are misspecified, $\widehat\tau^{ML}$, $\widehat\tau^{Prop}$, and $\widehat\tau^{Prog}$ are no longer consistent estimates of $\tau$.

\clearpage

\begin{table}
	\centering
	\caption{Performance of the estimation methods for recovering the average treatment effect. Six methods are compared, (1) Naïve inverse probability weighting (IPW), (2) Naïve outcome regression (OR), (3) maximum likelihood, (4) stratification based on propensity scores, (5) stratification based on prognostic scores, and (6) blocking. Absolute bias and root mean square error (RMSE) are reported, with all values multiplied by 100.}
    \rotatebox{270}{
	\begin{tabular}{ccc|cccccc}
		\hline
		& & & \multicolumn{6}{c}{Method} \\
		$(\eta_0, \eta_1)$ & Scenario & $N$ & Na\"ive IPW &  Na\"ive OR & ML & Prop & Prog & Block \\[0.1cm]
		\hline
		(0.1, 0.1) & (cor, cor) & 1000 & 0.553 (4.802) & 0.584 (4.629) & 0.040 (3.106) &  0.065 (3.338) & 0.399 (3.381) & 0.993 (3.391) \\[0.1cm]
		& & 2000 & 0.579 (3.449) & 0.568 (3.313) & 0.004 (2.117) & 0.016 (2.233) & 0.268 (2.291) & 0.570 (2.349) \\[0.1cm]
		& (cor, mis) & 1000 & 0.310 (4.842) & 0.228 (4.532) & 0.238 (3.097) & 0.056 (3.289) & 0.143 (3.246) & 0.044 (3.226) \\[0.1cm]
		& & 2000 & 0.529 (3.560) & 0.266 (3.336) & 0.201 (2.174) & 0.008 (2.284) & 0.157 (2.280) & 0.049 (2.289) \\[0.1cm]
		& (mis, cor) & 1000 & 0.352 (4.044) & 0.383 (4.025) & 0.031 (2.733) & 0.063 (2.879) & 0.075 (2.834) & 0.150 (2.913) \\[0.1cm]
		& & 2000 & 0.610 (2.900) & 0.636 (2.892) & 0.015 (1.926) & 0.025 (1.954) & 0.046 (1.977) & 0.106 (2.027) \\[0.1cm]
		& (mis, mis) & 1000 & 2.050 (4.557) & 2.014 (4.531) & 2.952 (4.023) & 2.540 (3.837) & 2.658 (3.896) & 1.860 (3.432) \\[0.1cm]
		& & 2000 & 2.273 (3.649) & 2.235 (3.613) & 3.106 (3.659) & 2.594 (3.318) & 2.782 (3.444) & 1.384 (2.457) \\[0.1cm]
		\hline
		(0.1, 0.2) & (cor, cor) & 1000 & 4.843 (7.066) & 4.910 (6.922) & 0.061 (3.274) & 0.050 (3.567) & 0.321 (3.517) & 0.877 (3.565) \\[0.1cm]
		& & 2000 & 4.523 (5.717) & 4.654 (5.754) & 0.051 (2.245) & 0.012 (2.334) & 0.207 (2.404) & 0.371 (2.431) \\[0.1cm]
		& (cor, mis) & 1000 & 4.608 (6.993) & 4.659 (6.733) & 0.271 (3.262) & 0.337 (3.541) & 0.465 (3.484) & 0.281 (3.397) \\[0.1cm]
		& & 2000 & 4.881 (6.076) & 4.765 (5.884) & 0.026 (2.302) & 0.004 (2.390) & 0.233 (2.381) & 0.108 (2.445) \\[0.1cm]
		& (mis, cor) & 1000 & 5.155 (6.475) & 5.148 (6.447) & 0.062 (2.881) & 0.137 (2.998) & 0.180 (2.978) & 0.062 (3.073) \\[0.1cm]
		& & 2000 & 4.922 (5.817) & 4.960 (5.846) & 0.086 (2.081) & 0.027 (2.139) & 0.021 (2.099) & 0.034 (2.208) \\[0.1cm]
		& (mis, mis) & 1000 & 2.334 (4.869) & 2.397 (4.891) & 3.370 (4.443) & 2.916 (4.198) & 3.006 (4.296) & 2.164 (3.743) \\[0.1cm]
		& & 2000 & 2.255 (3.724) & 2.288 (3.733) & 3.129 (3.725) & 2.663 (3.407) & 2.778 (3.490) & 1.403 (2.557) \\[0.1cm]
		\hline
	\end{tabular}}
	\label{table2}
\end{table}

\clearpage
\section{Data Example: Child Abuse and Adult Anger}
\label{sec6}
In this section, we apply the causal inference framework to the motivating example of our research, which examines the causal relationship between childhood abuse and adult anger. We consider a retrospective cohort study to examine the question, \textit{``Does child abuse by either parent increase a likelihood toward to adult anger?"}. This study focuses on the publicly available 1993-1994 sibling survey of the Wisconsin Longitudinal Study (WLS). The treatment is defined as the presence or absence of childhood abuse by either the father or mother, and the outcome is determined by a binary indicator of whether either parent exhibits a high anger score. See \cite{springer2007long, small2013case} for additional details regarding the WLS data.

\cite{springer2007long} indicated that the results might be affected by a tendency to under-reporting of abuse. Adults are likely not to report their childhood abuse even though there is any. With this information, we applied (1) ML, (2) propensity score stratification, (3) prognostic score stratification, and (4) blocking for the estimation of the ATE. The logistic outcome regression with the seven covariates without interaction terms is considered for the ML method. The same exposure model is used for propensity score stratification, whereas prognostic score stratification is based on the same outcome model. Ten strata are constructed by using the quantile values of the estimated score. A block size of 20 is used for the blocking method to build blocks. \cite{fergusson2000stability} asserted that a severe amount of false negative responses (approximately 50\%) exist when reporting childhood abuse, whereas false positive responses are absent. Based on this study, we can consider $0 \leq \eta_0(\xv), \eta_1(\xv) \leq \delta$ by letting $\delta \leq 0.5$ and compute the bounds according to $\delta$ as $\delta$ increases to 0.5.

\begin{figure}
\begin{center}
\centerline{\includegraphics[width=1\linewidth]{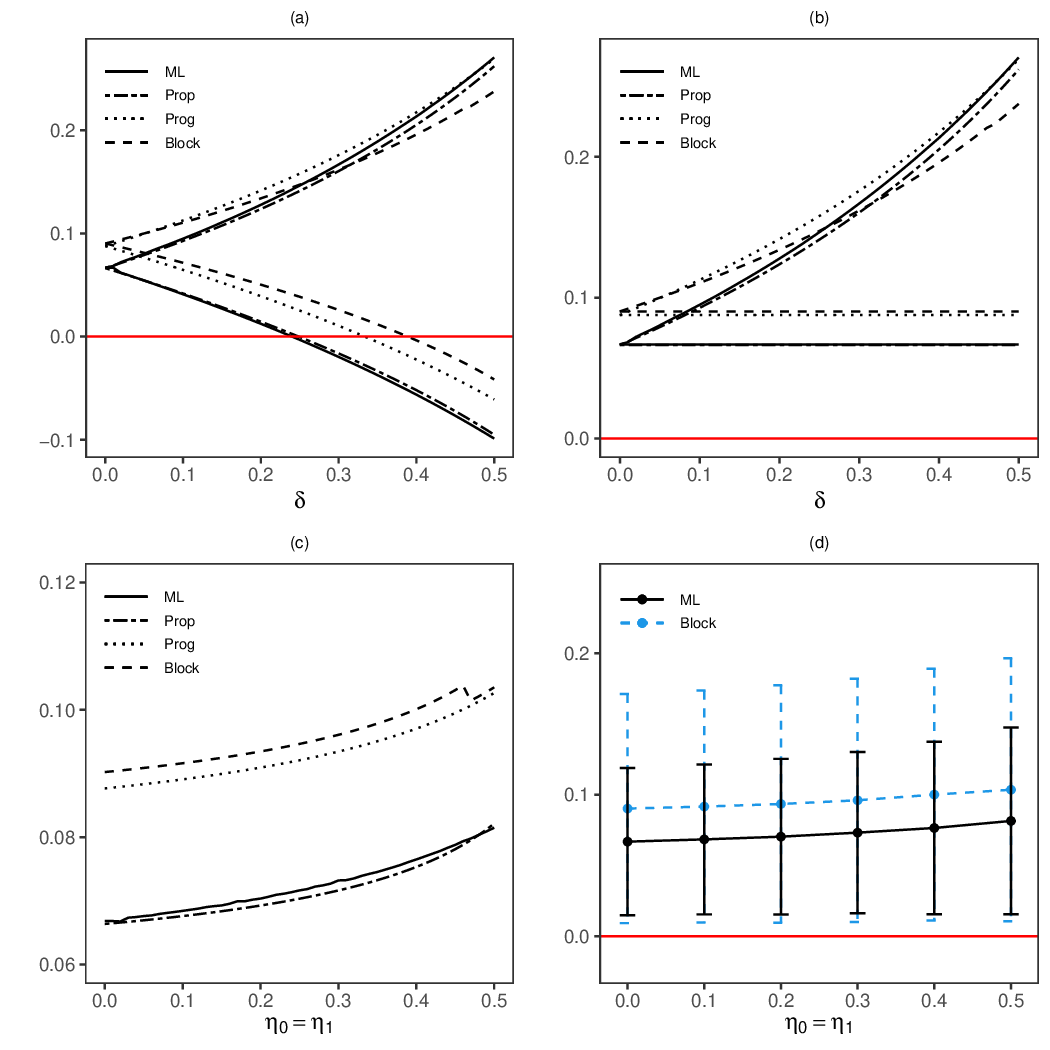}}
\end{center}
\caption{(a) Bounds of the average treatment effect (ATE) with $0 \leq\eta_0, \eta_1 \leq \delta$, (b) bounds of the ATE with $0 \leq\eta_0 \leq \eta_1 \leq \delta$, (c) point estimates of the ATE across the line of $\eta_0 = \eta_1$, and (d) point estimates of the ATE from maximum likelihood (ML) and blocking methods with 95\% bootstrapped confidence intervals.}
\label{figure1}
\end{figure}

As shown in Figure~\ref{figure1}(a), the bounds become wider as $\delta$ increases. All four bounds are above zero until $\delta = 0.22$. It is shown that they have a similar pattern, but the blocking method is the least sensitive to $\delta$. Since the prognostic score stratification is similar to the blocking, this figure may indicate that the propensity score model is misspecified. Moreover, \cite{deblinger2005understanding} stated that individuals who have high anger scores are more likely to experience recall bias when reporting childhood abuse experiences, that is, $\eta_0(\xv)\leq \eta_1(\xv)$. This allows us to narrow down the lower bounds of the ATE from Proposition~\ref{proposition2}, as presented in Figure~\ref{figure1}(b). Hence, we can conclude that childhood abuse has a causal effect on high anger scores in individuals, even in the presence of differential recall bias, based on this assumption without knowing $\eta_0(\xv), \eta_1(\xv)$.

\begin{table}
	\centering
	\caption{The effects of recall bias for six values of $\eta_0= \eta_1$. The estimates and 95\% bootstrap confidence intervals are displayed for the maximum likelihood (ML) and stratification methods}
	\begin{tabular}{ccccc}
		\hline
		& \multicolumn{4}{c}{Method}\\
		$(\eta_0, \eta_1)$ & ML & Prop & Prog & Block  \\
		\hline
		(0.0,0.0) & 0.067 ($\pm$ 0.052) & 0.066 ($\pm$ 0.035) & 0.088 ($\pm$ 0.055) & 0.090 ($\pm$ 0.081) \\
		(0.1,0.1) & 0.068 ($\pm$ 0.053) & 0.068 ($\pm$ 0.035) & 0.089 ($\pm$ 0.055) & 0.091 ($\pm$ 0.082) \\
		(0.2,0.2) & 0.070 ($\pm$ 0.055) & 0.069 ($\pm$ 0.035) & 0.091 ($\pm$ 0.056) & 0.093 ($\pm$ 0.084) \\
		(0.3,0.3) & 0.073 ($\pm$ 0.057) & 0.072 ($\pm$ 0.034) & 0.093 ($\pm$ 0.058) & 0.096 ($\pm$ 0.086) \\
		(0.4,0.4) & 0.076 ($\pm$ 0.061) & 0.075 ($\pm$ 0.033) & 0.097 ($\pm$ 0.059) & 0.100 ($\pm$ 0.089) \\
		(0.5,0.5) & 0.081 ($\pm$ 0.066) & 0.082 ($\pm$ 0.033) & 0.103 ($\pm$ 0.062) & 0.106 ($\pm$ 0.093) \\
		\hline
	\end{tabular}
	\label{table3}
\end{table}

We may further narrow down the bounds of the ATE if we make a stronger assumption. \cite{fergusson2000stability} suggested that the probabilities of recall bias may not differ significantly based on an adult's anger score. This suggests that recall bias may not be strongly related to an individual's level of anger. This allows us to assume $\eta_0(\xv) = \eta_1(\xv)$. \cite{robins1985early} pointed out that there is a minimal impact of reporters' demographic characteristics, such as sex, age, and social class, on recall bias, which further allows us to assume $\eta_0(\xv) = \eta_1(\xv) = \eta$. The setup of the range $0 \leq \eta_0 = \eta_1 \leq 0.5$ requires the strongest assumption, but it helps us to look at the results succinctly. The estimates for various values of $\eta_0$ and $\eta_1$ are shown in Table~\ref{table3}. For this case, the variance estimation can be accompanied so that we can provide confidence intervals for various $\delta$ values. Figure~\ref{figure1}(c) shows the estimates of the ATE across the line of $\eta_0=\eta_1$. All the estimates increase as $\eta_0 = \eta_1$ increases. Furthermore, the 95\% CIs of all methods do not contain 0. In Figure~\ref{figure1}(d), we particularly focus on the results of the ML and blocking estimators when $\eta_0=\eta_1$. Even though the confidence interval of the blocking estimator is broader than that of the ML estimator, possibly due to the fact that it requires weak assumptions, the confidence interval still stays above 0. These results imply that the under-reporting issue does not alter the initial conclusion; on the contrary, it strengthens the conclusion that there is significant evidence that child abuse increases the adult anger score.

\begin{figure}
\begin{center}
\centerline{\includegraphics[width=1\linewidth]{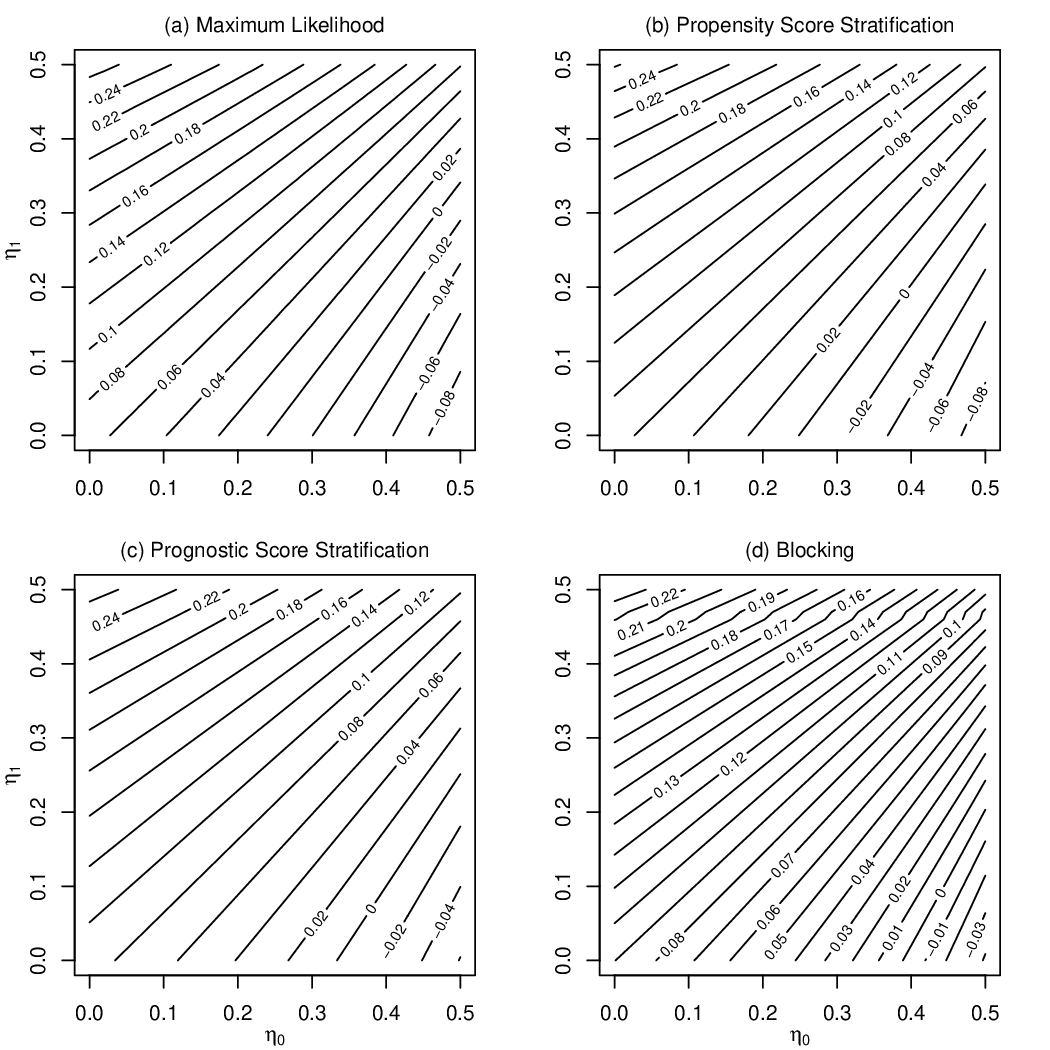}}
\end{center}
\caption{Contour plots for the values of $(\eta_0, \eta_1)$ using (a) maximum likelihood, (b) propensity score stratification, (c) prognostic score stratification, and (d) blocking methods in the region $0 \leq \eta_0, \eta_1 \leq 0.5$.
}
\label{figure2}
\end{figure}

We also conduct a sensitivity analysis of recall bias with parameters $0 \leq \eta_0, \eta_1 \leq 0.5$. Figure~\ref{figure2} shows contour plots of the estimated ATEs for the values of $\eta_0$ and $\eta_1$ in this region. This figure reveals that most of the estimates are above 0. Especially for the blocking method, estimates are below 0 only for a small region of $\eta_0 \geq \frac{1}{2}\eta_1 + 0.4$ for $0 \leq \eta_1 \leq 0.2$.

\clearpage

\section{Discussion}\label{sec7}

In this paper, we derived the ATE bounds and incorporated knowledge from prior research in order to narrow the bounds. Also, we proposed several approaches to estimate the bounds. Most of discussion were focused on the ATE, but the same argument can be applied to different measures such as average treatment effect on the treated (ATT), risk ratio, or odds ratio. In Web Appendix, we include more detailed discussions about these measures.

Also, another difficulty in applying the stratification methods proposed arises in computation. For instance, in some cases, both $a_i^*$ and $b_i^*$ are zero. To avoid any computational issues, we also propose a nearest-neighbor combination method to improve the stability of the estimates. This method is discussed in Web Appendix.

Finally, one limitation is that we cannot assess the covariate balance before adjustment since the exposure variable is misclassified. We can consider an indirect approach to check the covariate balance under the constant recall bias assumption. If stratification is successful, we may assume that the covariate distributions between the treated and control groups are equal, at least asymptotically. Also, if the magnitude of recall bias is independent of the covariates, we are able to assess the balance with a biased treatment $Z^*$. First, we calculate the number of treated and control units with bias correction such as $a_i^*/(1-\eta_1) + b_i^*/(1-\eta_0)$ and $c_i^* + d_i^* - a_i^*(\eta_1/(1-\eta_1))- b_i^*(\eta_0/(1-\eta_0))$ respectively. Second, under the assumption that the covariate vectors within each stratum are similar, we can compute the average covariate vectors for each stratum. Every individual in the same stratum shares the same average vector, which is not restrictive since we are going to compare the weighted means after all. Finally, we compare the absolute standardized mean difference between the treated and control groups using corrected weights based on this assumption. This new technique is also discussed in Web Appendix.

\section*{Acknowledgments}
This work was supported by NIH grants (R01ES026217, R01MD012769, R01ES028033, 1R01ES030616, 1R01AG066793-01R01, 1R01ES029950, R01ES028033-S1), Alfred P. Sloan Foundation (G-2020-13946), Vice Provost for Research at Harvard University (Climate Change Solutions Fund), the New Faculty Startup Fund from Seoul National University, the National Research Foundation of Korea (NRF) grant funded by the Korea government (MSIT) (2021R1C1C1012750), and the Global-LAMP Program of the National Research Foundation of Korea (NRF) grant funded by the Ministry of Education (No. RS-2023-00301976).

\section*{Supplementary Materials}
Supplementary materials containing the Web Appendices are available online. The R codes are available at \url{https://github.com/suhwanbong121/recall_bias_observational_study}.

\bibliographystyle{apalike}

\begin{thebibliography}{}

\bibitem[Armstrong, 1985]{armstrong1985measurement}
Armstrong, B. (1985).
\newblock Measurement error in the generalised linear model.
\newblock {\em Communications in Statistics-Simulation and Computation},
  14(3):529--544.

\bibitem[Babanezhad et~al., 2010]{babanezhad2010comparison}
Babanezhad, M., Vansteelandt, S., and Goetghebeur, E. (2010).
\newblock Comparison of causal effect estimators under exposure
  misclassification.
\newblock {\em Journal of Statistical Planning and Inference},
  140(5):1306--1319.

\bibitem[Braun et~al., 2017]{braun2017propensity}
Braun, D., Gorfine, M., Parmigiani, G., Arvold, N.~D., Dominici, F., and
  Zigler, C. (2017).
\newblock Propensity scores with misclassified treatment assignment: a
  likelihood-based adjustment.
\newblock {\em Biostatistics}, 18(4):695--710.

\bibitem[Bross, 1954]{bross1954misclassification}
Bross, I. (1954).
\newblock Misclassification in 2 x 2 tables.
\newblock {\em Biometrics}, 10(4):478--486.

\bibitem[Carroll et~al., 1985]{carroll1985comparison}
Carroll, R.~J., Gallo, P., and Gleser, L.~J. (1985).
\newblock Comparison of least squares and errors-in-variables regression, with
  special reference to randomized analysis of covariance.
\newblock {\em Journal of the American Statistical Association},
  80(392):929--932.

\bibitem[Carroll et~al., 1995]{carroll1995measurement}
Carroll, R.~J., Ruppert, D., and Stefanski, L.~A. (1995).
\newblock {\em Measurement error in nonlinear models}.
\newblock New York: Chapman and Hall.

\bibitem[Carroll and Stefanski, 1990]{carroll1990approximate}
Carroll, R.~J. and Stefanski, L.~A. (1990).
\newblock Approximate quasi-likelihood estimation in models with surrogate
  predictors.
\newblock {\em Journal of the American Statistical Association},
  85(411):652--663.

\bibitem[Cochran, 1968]{cochran1968errors}
Cochran, W.~G. (1968).
\newblock Errors of measurement in statistics.
\newblock {\em Technometrics}, 10(4):637--666.

\bibitem[Deblinger and Runyon, 2005]{deblinger2005understanding}
Deblinger, E. and Runyon, M.~K. (2005).
\newblock Understanding and treating feelings of shame in children who have
  experienced maltreatment.
\newblock {\em Child Maltreatment}, 10(4):364--376.

\bibitem[Fergusson et~al., 2000]{fergusson2000stability}
Fergusson, D.~M., Horwood, L.~J., and Woodward, L.~J. (2000).
\newblock The stability of child abuse reports: a longitudinal study of the
  reporting behaviour of young adults.
\newblock {\em Psychological Medicine}, 30(3):529--544.

\bibitem[Fuller, 1980]{fuller1980properties}
Fuller, W.~A. (1980).
\newblock Properties of some estimators for the errors-in-variables model.
\newblock {\em The Annals of Statistics}, 8(2):407--422.

\bibitem[Gleser, 1990]{gleser1990improvements}
Gleser, L.~J. (1990).
\newblock Improvements of the naive approach to estimation in nonlinear
  errors-in-variables regression models.
\newblock {\em Contemporary Mathematics}, 112:99--114.

\bibitem[Gravel and Platt, 2018]{gravel2018weighted}
Gravel, C.~A. and Platt, R.~W. (2018).
\newblock Weighted estimation for confounded binary outcomes subject to
  misclassification.
\newblock {\em Statistics in Medicine}, 37(3):425--436.

\bibitem[Hansen, 2008]{hansen2008prognostic}
Hansen, B.~B. (2008).
\newblock The prognostic analogue of the propensity score.
\newblock {\em Biometrika}, 95(2):481--488.

\bibitem[Imai and Yamamoto, 2010]{imai2010causal}
Imai, K. and Yamamoto, T. (2010).
\newblock Causal inference with differential measurement error: Nonparametric
  identification and sensitivity analysis.
\newblock {\em American Journal of Political Science}, 54(2):543--560.

\bibitem[Karmakar et~al., 2021]{karmakar2021reinforced}
Karmakar, B., Small, D.~S., and Rosenbaum, P.~R. (2021).
\newblock Reinforced designs: Multiple instruments plus control groups as
  evidence factors in an observational study of the effectiveness of catholic
  schools.
\newblock {\em Journal of the American Statistical Association},
  116(533):82--92.

\bibitem[Lindley, 1953]{lindley1953estimation}
Lindley, D. (1953).
\newblock Estimation of a functional relationship.
\newblock {\em Biometrika}, 40(1/2):47--49.

\bibitem[Lockwood and McCaffrey, 2016]{lockwood2016matching}
Lockwood, J.~R. and McCaffrey, D.~F. (2016).
\newblock Matching and weighting with functions of error-prone covariates for
  causal inference.
\newblock {\em Journal of the American Statistical Association},
  111(516):1831--1839.

\bibitem[Lord, 1960]{lord1960large}
Lord, F.~M. (1960).
\newblock Large-sample covariance analysis when the control variable is
  fallible.
\newblock {\em Journal of the American Statistical Association},
  55(290):307--321.

\bibitem[McCaffrey et~al., 2013]{mccaffrey2013inverse}
McCaffrey, D.~F., Lockwood, J.~R., and Setodji, C.~M. (2013).
\newblock Inverse probability weighting with error-prone covariates.
\newblock {\em Biometrika}, 100(3):671--680.

\bibitem[Raphael, 1987]{raphael1987recall}
Raphael, K. (1987).
\newblock Recall bias: a proposal for assessment and control.
\newblock {\em International Journal of Epidemiology}, 16(2):167--170.

\bibitem[Robins et~al., 1985]{robins1985early}
Robins, L.~N., Schoenberg, S.~P., Holmes, S.~J., Ratcliff, K.~S., Benham, A.,
  and Works, J. (1985).
\newblock Early home environment and retrospective recall: A test for
  concordance between siblings with and without psychiatric disorders.
\newblock {\em American Journal of Orthopsychiatry}, 55(1):27--41.

\bibitem[Rosenbaum and Rubin, 1983]{rosenbaum1983central}
Rosenbaum, P.~R. and Rubin, D.~B. (1983).
\newblock The central role of the propensity score in observational studies for
  causal effects.
\newblock {\em Biometrika}, 70(1):41--55.

\bibitem[Rosner et~al., 1990]{rosner1990correction}
Rosner, B., Spiegelman, D., and Willett, W.~C. (1990).
\newblock Correction of logistic regression relative risk estimates and
  confidence intervals for measurement error: the case of multiple covariates
  measured with error.
\newblock {\em American Journal of Epidemiology}, 132(4):734--745.

\bibitem[Rosner et~al., 1989]{rosner1989correction}
Rosner, B., Willett, W., and Spiegelman, D. (1989).
\newblock Correction of logistic regression relative risk estimates and
  confidence intervals for systematic within-person measurement error.
\newblock {\em Statistics in Medicine}, 8(9):1051--1069.

\bibitem[Rothman, 2012]{rothman2012epidemiology}
Rothman, K.~J. (2012).
\newblock {\em Epidemiology: an introduction}.
\newblock New York: Oxford University Press.

\bibitem[Rothman et~al., 2008]{rothman2008modern}
Rothman, K.~J., Greenland, S., Lash, T.~L., et~al. (2008).
\newblock {\em Modern epidemiology}, volume~3.
\newblock Wolters Kluwer Health/Lippincott Williams \& Wilkins Philadelphia.

\bibitem[Rubin, 1974]{rubin1974estimating}
Rubin, D.~B. (1974).
\newblock Estimating causal effects of treatments in randomized and
  nonrandomized studies.
\newblock {\em Journal of Educational Psychology}, 66(5):688--701.

\bibitem[Rubin, 1980]{rubin1980randomization}
Rubin, D.~B. (1980).
\newblock Randomization analysis of experimental data: The fisher randomization
  test comment.
\newblock {\em Journal of the American Statistical Association},
  75(371):591--593.

\bibitem[Small et~al., 2013]{small2013case}
Small, D.~S., Cheng, J., Halloran, M.~E., and Rosenbaum, P.~R. (2013).
\newblock Case definition and design sensitivity.
\newblock {\em Journal of the American Statistical Association},
  108(504):1457--1468.

\bibitem[Springer et~al., 2007]{springer2007long}
Springer, K.~W., Sheridan, J., Kuo, D., and Carnes, M. (2007).
\newblock Long-term physical and mental health consequences of childhood
  physical abuse: Results from a large population-based sample of men and
  women.
\newblock {\em Child Abuse \& Neglect}, 31(5):517--530.

\bibitem[Stefanski and Carroll, 1985]{stefanski1985covariate}
Stefanski, L.~A. and Carroll, R.~J. (1985).
\newblock Covariate measurement error in logistic regression.
\newblock {\em The Annals of Statistics}, 14(3):1335--1351.

\end{thebibliography}

\end{document}